\documentclass[letterpaper,english,aps,prl,twocolumn,floatfix,showpacs,amsfonts,amssymb,superscriptaddress]{revtex4}

\usepackage[T1]{fontenc}
\usepackage[latin1]{inputenc}
\usepackage{graphics}
\usepackage{amssymb}
\usepackage{overpic}

\newcommand{\beq}{\begin{equation}}
\newcommand{\eeq}{\end{equation}}
\newcommand{\bea}{\begin{eqnarray}}
\newcommand{\eea}{\end{eqnarray}}

\newcommand{\rmd}{{\rm d}}
\newcommand{\rmi}{{\rm i}}

\begin{document}

\title{Magnetic and Quasiparticle Spectra of an itinerant $J_1-J_2$ Model for Iron Pnictides}

\author{C.~M.~S.~da Concei\c{c}\~ao}

\affiliation{Departamento de F\'isica Te\'orica, Universidade do Estado do Rio de Janeiro, Rio de Janeiro, RJ 20550-013, Brasil}

\author{M.~B.~Silva~Neto}

\affiliation{Instituto de F\' isica, Universidade Federal do
Rio de Janeiro, Caixa Postal 68528, Brasil}

\author{E.~C.~Marino}

\affiliation{Instituto de F\' isica, Universidade Federal do
Rio de Janeiro, Caixa Postal 68528, Brasil}

\begin{abstract}

We calculate the magnetic and quasiparticle excitation spectra of an itinerant $J_1-J_2$ 
model for iron pnictides. In addition to an acoustic spin-wave branch, the magnetic spectrum 
has a second, optical branch, resulting from the coupled four-sublattice magnetic structure. 
The spin-wave velocity has also a planar directional anisotropy, due to the collinear/striped 
antiferromagnetism. Within the magnetically ordered phase, the quasiparticle spectrum is composed 
of two Dirac cones, resulting from the folding of the magnetic Brillouin zone. We discuss the 
relevance of our findings to the understanding of both neutron scattering and photoemission 
spectroscopy results for SrFe$_{2}$As$_{2}$.

\end{abstract}


\maketitle

Since the discovery of superconductivity above $50$ K in RFeAsO$_{1-x}$F$_x$
(R $=$ La, Ce, Sm, etc.) \cite{SC-1111-Pnictides}, the term {\it high temperature 
superconductivity} can no longer be associated exclusively to cuprates. 
The new iron-based pnictide compounds, including the MFe$_{2}$As$_{2}$ 
(M $=$ Ba, Sr, etc.) family \cite{SC-122-Pnictides}, exhibit, just like cuprates, 
a layered antiferromagnetic (AF) structure in the parent compound, which 
gives room to superconductivity upon doping \cite{Mike-Norman}. One 
important difference, however, is related to the transport and optical properties 
of their parent compounds. While cuprates are Mott antiferromagnetic 
insulators, pnictides are metals, albeit not very good ones, with a very rich 
multi-band Fermi surface \cite{FS-BaKFe2As2}. Despite this difference, it 
is believed that, in both cases, strong correlations of strength $U$ among 
transition metal electrons, from Cu in the case of cuprates and from Fe in 
the case of pnictides, play a very important role \cite{Abrahams}. In fact, 
it has been argued that, at zero temperature, $T=0$, these two systems 
could be represented as very closely, but at oposite sides, to a Mott metal 
insulator transition, in the $T\times U$ phase diagram \cite{Abrahams}. 
This would explain, for example, the experimentally observed strong 
renormalization of the coherent quasiparticle spectral weight at the Fermi 
level in pnictides \cite{Hall-Spectroscopy}, which would be shifted to form 
incoherent lower and upper Hubbard bands, just like in prototypical Mott 
insulators such as cuprates. 

The incoherent part of the spectrum in iron pnictides can be well described 
in terms of localized moments at the Fe positions, interacting via a 
superexchange $J_{1}$ between nearest Fe neighbors, Fe$-$Fe, and 
a second superexchange $J_{2}$ between As bridged next-to-nearest 
Fe neighbors, Fe$-$As$-$Fe \cite{Superexchange}. Alternatively, an 
itinerant description, in terms of coherent multi-band electrons and holes, 
Fermi surface nesting, and spin-density-wave (SDW) instability, can also 
be used for the understanding of both the antiferromagnetic and 
superconducting states in iron pnictides 
\cite{Tesanovic,Chubukov}. In this letter, we shall adopt a combined 
description, merging these two degrees of freedom, in which local 
magnetic moments are coupled, via Hund's exchange interaction 
\cite{Philip-Phillips}, to itinerant quasiparticles, to be 
described by a minimal two-band model \cite{Raghu}. 

Let us start by analyzing the magnetic excitation spectrum. 
One important difference between having $S=1/2$, as in cuprates, or 
$S=1$ (or higher), as in pnictides, is the possibility of single ion anisotropy, 
in the later case, which arises from relativistic corrections and the spin
orbit coupling. The simplest model that captures all the essential local
moment physics described in the previous paragraphs is the 
extended $J_{1}-J_{2}$ spin-Hamiltonian
\beq
\hat{H}=J_{1}\sum_{\langle i,j\rangle}{\bf \hat{S}}_{i}\cdot {\bf \hat{S}}_{j}+
J_{2}\sum_{\langle\langle i,j\rangle \rangle}{\bf \hat{S}}_{i}\cdot {\bf \hat{S}}_{j}-
K\sum_{i}(\hat{S}_i^z)^2,
\label{J1-J2-Hamiltonian}
\eeq
where $J_{1}>0$ and $J_{2}>0$ are, respectively, the antiferromagnetic
superexchanges between nearest-neighbors, $\langle i,j\rangle$, and 
next-to-nearest neighbors, $\langle \langle i,j\rangle \rangle$, spins 
$\hat{\bf S}_i$ on a two dimensional square lattice, and $K>0$ is the 
single ion anisotropy coupling constant. 

The existence of two superexchanges $J_{1}$ and $J_{2}$ in the spin 
Hamiltonian (\ref{J1-J2-Hamiltonian}) renders the antiferromagnetism 
collinear, with wave vectors at $(\pi,0)$ and/or $(0,\pi)$ \cite{Dai}. This
is in agreement with inelastic neutron scattering in
(Sr,Ba)Fe$_{2}$As$_{2}$ \cite{Dai,Ewings}, which exhibit peaks
at the AF zone center $Q=(1,0,1)$, for energies between
$5-15$ meV. This situation is markedly different from a N\' eel 
ordered state with wave vector at $(\pi,\pi)$, as found in cuprates, where 
a single superexchange $J$ between Cu$-$O$-$Cu is present. As 
discussed in the literature, such wave vector degeneracy in the classical 
ground state of the Hamiltonian (\ref{J1-J2-Hamiltonian}) gives rise to 
an extra Ising symmetry  \cite{Chandra,Sachdev,Kivelson}, which is 
broken at a different temperature than the AF ordering one. The 
hierarchy of symmetry breaking as $T$ is lowered is: paramagnetic 
(PM)/Ising symmetric, PM/Ising broken, and AF/Ising broken phases
\cite{Sachdev}.

Previous linear spin wave studies of the $J_{1}-J_{2}$ model with $K=0$ 
have considered, as a starting point, a classical ground state with a {\it two 
sublattice structure}, and have obtained that the quantized magnetic 
spectrum would be composed by a single gapless spin-wave branch 
\cite{Carlson}, with $\hbar\omega({\bf k})\sim |{\bf k}|$ for small wave 
vector ${\bf k}$. The existence of such gapless mode reflects the full 
rotational invariance of the spin Hamiltonian (\ref{J1-J2-Hamiltonian}), 
when $K=0$, and is in agreement with Goldstone's theorem. We shall 
refer to this branch as the {\it acoustic branch}. On the other hand, 
inelastic neutron scattering revealed the gapped nature of the magnetic 
spectrum \cite{Dai,Ewings}, indicating that the single-ion anisotropy
parameter $K\neq 0$. However, the absence of a clean step in the neutron 
integrated intensity (expected for the case of a single gap), makes room 
for the existence of two or more magnetic branches \cite{Dai,Ewings}. 
As we shall soon see, the actual {\it four sublattice structure} of the AF/Ising 
broken classical ground state, with two coupled, interpenetrating N\' eel 
ordered states, gives rise to a second, optical spin-wave branch (already 
for $K=0$). Below we will consider $K\neq 0$, when 
rotational invariance is broken and all branches become 
gapped. 




In order to understand the origin of the optical mode in the $J_{1}-J_{2}$
Hamiltonian (\ref{J1-J2-Hamiltonian}) it suffices to consider the isotropic 
case, $K=0$. For $J_{2}\simeq 2J_{1}$, as the experiments suggest, the 
ground state is a collinear antiferromagnet, composed by two coupled, 
interpenetrating N\' eel ordered states
We span spins in a coherent basis for each of these N\' eel states, 
which we label as $A$ and $B$, and we write
$
{\mathbf{\hat{S}}}^{A}_{i}=S\mathbf{\Omega}^{A}_{i}=
S\left[e^{\rmi{\mathbf{Q}}\cdot{\mathbf{x}}_i}{\mathbf{n}}_{A}({\mathbf{x}}_i)\sqrt{1-
\left(\frac{{\mathbf{L}}_{A}}{\bar{s}}\right)^2}+\frac{{\mathbf{L}}_{A}}{\bar{s}}\right]$, 
and
$
{\mathbf{\hat{S}}}^{B}_{i}=S\mathbf{\Omega}^{B}_{i}=
S\left[e^{\rmi{\mathbf{Q}}\cdot{\mathbf{x}}_i}{\mathbf{n}}_{B}({\mathbf{x}}_i)\sqrt{1-
\left(\frac{\mathbf{L}_{B}}{\bar{s}}\right)^2}+\frac{{\mathbf{L}}_{B}}{\bar{s}}\right]
$,
where ${\bf Q}=(\pi/a,\pi/a)$ is the ordering wave vector for each N\' eel state, 
${\bf n}_{A,B}$ and ${\bf L}_{A,B}$ are, respectively, the staggered and uniform 
components of the spins belonging to the two states, and $\bar{s}=S/a^{d}$ 
is the density of spin in the unit cell. We use the constraint 
${\mathbf{n}}_{A,B}^2=1$.

After integrating out the uniform (fast) components ${\mathbf{L}}_{A}$ and ${\mathbf{L}}_{B}$
of the spins we arrive at the action of the nonlinear sigma model  for the case $K=0$ (the
detailed derivation of this model will appear elsewhere \cite{Next}). This action describes the 
low-energy, long-wavelength fluctuations of the staggered (slow) order parameter ${\bf n}_{A,B}$ 
(as usual we use $\beta=1/k_B T$ and $\int=\int_0^{\hbar\beta}\rmd\tau\int\rmd^2{\bf x}$)
%
{\small \bea
{\cal S}&=&\frac{\rho_{s}}{2\hbar}\int
\left\{\left(|\nabla{\bf n}_{A}|^2+|\nabla{\bf n}_{B}|^2\right)+
\frac{1}{c_{0}^2}\left(|\partial_\tau{\bf n}_{A}|^2+|\partial_\tau{\bf n}_{B}|^2\right)\right.\nonumber\\
&+&\left.\gamma\left({\bf n}_{A} \cdot\partial_x\partial_y {\bf n}_{B}+{\bf n}_{B} \cdot\partial_x\partial_y {\bf n}_{A}\right)+
\eta\;{\bf n}_{A}\cdot{\bf n}_{B}\right.\nonumber\\
&+&\left.\rmi b\left[{\bf n}_{A}\cdot({\bf n}_{B}\times\partial_\tau {\bf n}_{B})+
{\bf n}_{B}\cdot({\bf n}_{A}\times\partial_\tau {\bf n}_{A})\right]\right.\nonumber\\
&-&\left.\frac{1}{c_{1}^2}\left[({\bf n}_{A}\cdot{\bf n}_{B})(\partial_\tau{\bf n}_{A}\cdot\partial_\tau{\bf n}_{B})
-({\bf n}_{A}\cdot\partial_\tau{\bf n}_{B})({\bf n}_{B}\cdot\partial_\tau{\bf
  n}_{A})\right]\right\}.
\label{Action}
\eea}
%
The first line describes the low-energy, long-wavelength fluctuations of the 
two order parameters ${\bf n}_A, {\bf n}_B$, independently. The second line 
contains their coupling, through $\gamma$, within the AF/Ising broken phase, 
where the term ${\bf n}_{A}\cdot{\bf n}_{B}$ is allowed (this model was derived 
for the $(\pi,0)$ magnetic configuration, which breaks explicitly the Ising symmetry).
For the Ising symmetric part of the phase diagram, such term would have been 
absent, but integration over fluctuations would give rise, instead, to a term like 
$({\bf n}_{A}\cdot{\bf n}_{B})^2$, which is Ising invariant \cite{Chandra}. The third 
line contains a dynamical term that describes the Bloch precession of the staggered 
moments in a given N\' eel state (say A) around the moments of the other (say B), 
and it resembles the coupling to a magnetic field. Finally, the fourth line contains 
interacting dynamical terms that also arise from the integration over ${\bf L}_{A,B}$. 

All couplings in (\ref{Action}) are expressed in terms of the original parameters. 
The spin stiffness is $\rho_{s}=2J_{2}S^{2}a^{2-d}$, the spin-wave velocity equals 
${c_{0}=2\sqrt{2}S a}\sqrt{4J_{2}^{2}-J_{1}^{2}}/\hbar$, we will also have 
a contribution from $c_{1}=c_{0}(J_2/J_1)$, while the couplings between 
${\bf n}_{A}$ and ${\bf n}_{B}$ are given by
$\gamma=\frac{2J_{1}}{J_{2}}\left(1+\frac{J_{1}^{2}/4}{4J_{2}^{2}-J_{1}^{2}}\right)$,
$\eta=\frac{2}{a^{2}}\left(\frac{J_{1}}{J_{2}}\right)\frac{J_{1}^{2}}{4J_{2}^{2}-J_{1}^{2}}$,
and finally $b=\frac{2\hbar J_{1}}{S^{2}}\frac{1}{(4J_{2}^{2}-J_{1}^{2})}$.
The bare (unrenormalized) values of the parameters described above are, as 
expected, higher than the measured values. For example, for $a=5.695$ {\AA}, 
$J_1=20$ meV,and $J_{2}=40$ meV, we find $\hbar c_0=1.2$ eV {\AA}, while the 
typical values are actually around $0.25$ eV {\AA}. The theory described by Eq. 
(\ref{Action}) is highly interacting and a full renormalization procedure is required 
to reproduce the actual values of all couplings and the smallness of the Fe 
magnetic moment \cite{Next}. For the purposes of comparison with 
experiments in this letter, we shall use the already established values for some 
of these constants.

\begin{figure}
\includegraphics[scale=.3]{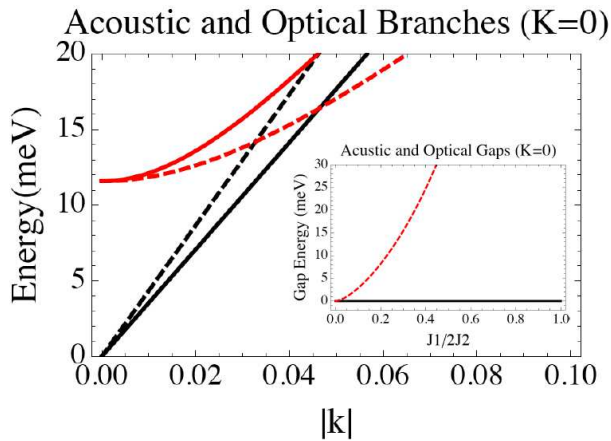}
\caption{(Color online): Acoustic (lower, black) and Optical (upper, red) spin-wave
branches for the isotropic case, $K=0$, as a function of $|k|$, for two different directions, 
$k_y=0$ (solid lines) and $k_x=k_y$ (dashed lines). Inset: Acoustic and Optical gaps
as a function of $J_{1}/2J_{2}$. Notice that while the acoustic gap is always 
zero, as imposed by Goldstone's theorem, the optical gap hardens as $J_{1}$ 
increases, diverging for $J_1=2J_2$. }
\label{Fig-Acoustical-Optical}
\end{figure}

To find the magnetic excitation spectrum we look at the poles of the staggered spectral 
function, ${\cal A}({\bf k},\omega)$. Within the Green's function formalism, these are 
obtained from the imaginary part of the retarded Green's function, $G_{ret}({\bf k},\omega)$, 
for transverse staggered fluctuations ${\cal A}({\bf k},\omega)=-(1/\pi)\lim_{\delta\rightarrow 0}
{\cal I}m[G_{ret}({\bf k},\rmi\omega_n\rightarrow\omega+\rmi\delta)]$, where 
$\omega_n=2\pi n/\beta$ are the Matsubara frequencies.
The first set of poles corresponds to
\bea
\hbar\omega({\bf k})=\hbar\, c_{SW}\sqrt{{\bf k}^2+\gamma k_x k_y},
\eea
where
$\frac{1}{c_{SW}^2}=\left(\frac{1}{c_{0}^{2}}-\frac{\sigma^{2}}{2c_{1}^{2}}\right)$.
This is the usual acoustic spin-wave branch also found in linear spin wave theory
\cite{Carlson}, which is gapless at the zone center, $\hbar\omega(0)=0$, in 
agreement with Goldstone's theorem. Notice that the observable spin-wave 
velocity $c_{SW}$ {\it increases} by a factor $\sqrt{2+\gamma}$ as one moves 
from the $k_y=0$ (or $k_x=0$) line to the $k_x=k_y$ direction (along the
spin stripes), showing that the spin-wave dispersion has a {\it planar directional 
anisotropy}, see Fig.\ \ref{Fig-Acoustical-Optical}, a consequence of the collinear 
character of the antiferromagnetic order. 
The second set of poles corresponds to
\beq
\hbar\Omega({\bf k})=
\hbar\, c_{op}\sqrt{{\bf k}^2-\gamma k_x k_y+\eta},
\eeq
where
$\frac{1}{ c_{op}^2}=\left(\frac{1}{c_{0}^{2}}+\frac{\sigma^2}{2c_{1}^{2}}\right)$.
Here, instead, we find an optical gap, $\hbar\Omega(0)=\Delta_{op}\neq 0$, 
at the antiferromagnetic zone center $(1,0,1)$, given by
\beq
\Delta_{op}=\hbar\, c_{op}\sqrt{\eta},
\eeq
which corresponds to long lived optical excitations. Notice now that the optical 
spin-wave velocity along the $k_x=k_y$ line {\it decreases} with respect to the
$k_y=0$ (or $k_x=0$) directions, see Fig.\ \ref{Fig-Acoustical-Optical}, exactly 
the opposite from the case for the acoustic branch. The optical gap hardens as 
$J_1$ increases, see inset in Fig.\ \ref{Fig-Acoustical-Optical}, and eventually 
diverges at $J_{1}=2J_{2}$, when the classical configuration changes and the 
action (\ref{Action}) has to be modified. 

After including a single ion anisotropy (SIA) term at the Fe ions, $K\neq 0$, we have 
extended the derivation of the nonlinear sigma model accordingly (details will be given 
elsewhere \cite{Next}) and we have recalculated the magnetic excitation spectrum for 
$K\neq 0$ \cite{Next}. The rather lengthy expressions for the dispersions of the two 
branches for arbitrary $K$ reduce, in the limit $K\ll J_2$, to
\bea
\hbar\omega({\bf k})&=&\hbar\, c_{SW}\sqrt{{\bf k}^2+\gamma k_x k_y+\Delta_{SIA}^2},
\label{Lower-Branch}
\\
\hbar\Omega({\bf k})&=&\hbar\, c_{op}\sqrt{{\bf k}^2-\gamma k_x k_y+\Delta_{SIA}^2+\eta}.
\label{Upper-Branch}
\eea
As expected, the immediate effect of a SIA term is to produce a gap
for all branches in the spectrum
\beq
\Delta_{SIA}^2=\frac{(S-1/2)}{S a^{2}}\left(\frac{K}{J_2}\right)
\left[1+\frac{1}{8}\frac{J_1^2}{(4J_2^2-J_1^2)}\right],
\eeq
which only exists for high spin systems, being absent when 
$S=1/2$, like in the case of cuprates. 

\begin{figure}
\includegraphics[scale=.3]{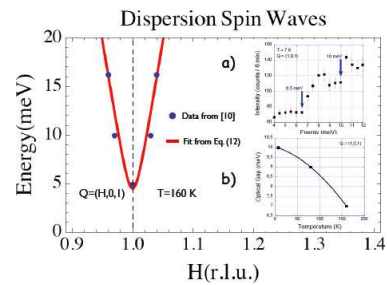}
\caption{(Color online): Spin-wave dispersion for  $Q=(H,0,1)$ and
at $T=160$ K. Experimental data from Ref. \cite{Dai}. Inset a): integrated
intensity at $T=7$ K, as a function of energy, exhibiting a two-step
profile consistent with the existence of two gapped branches in the 
magnetic spectrum, Eqs. (\ref{Lower-Branch}) and (\ref{Upper-Branch}).
Inset b): Temperature dependence of the optical gap, according to 
analysis of the data reported in Ref. \cite{Dai}.}
\label{Fig-Theory-Experiment}
\end{figure}

We can now analyze recent inelastic neutron scattering results for spin excitations 
in SrFe$_{2}$As$_{2}$ \cite{Dai}. The spin-wave dispersion shown in 
Fig.\ \ref{Fig-Theory-Experiment}, for scans along $(H,0,1)$ at $T=160$ K,  
can be well described by the lower branch of the spectrum, $\hbar\omega({\bf k})$, 
along the $k_x=k_y$ line  (we use a coordinate system rotated by $45^\circ$ with 
respect to the original unit cell). This branch is gapped solely by the SIA and for 
$T=7$ K the gap can be seen at $6.5$ meV, see the inset $a)$ of 
Fig.\ \ref{Fig-Theory-Experiment}. However, as discussed in the introduction, 
the broadened integrated intensity of the unpolarized neutrons, observed for 
either SrFe$_{2}$As$_{2}$ \cite{Dai} and BaFe$_{2}$As$_{2}$ \cite{Ewings}, 
makes room for the existence of two or more gaps in the magnetic spectrum. 
In fact, the integrated intensity shown in the inset $a)$ of 
Fig.\ \ref{Fig-Theory-Experiment} shows a two-step profile, indicating the presence 
a second magnon gap at $10$ meV. The temperature dependence of the optical 
gap, following a similar analysis of the data in \cite{Dai} for $T=7,80,160$ K, 
is shown in the inset $b)$ of Fig.\ \ref{Fig-Theory-Experiment}.

\begin{figure}
\includegraphics[scale=.3]{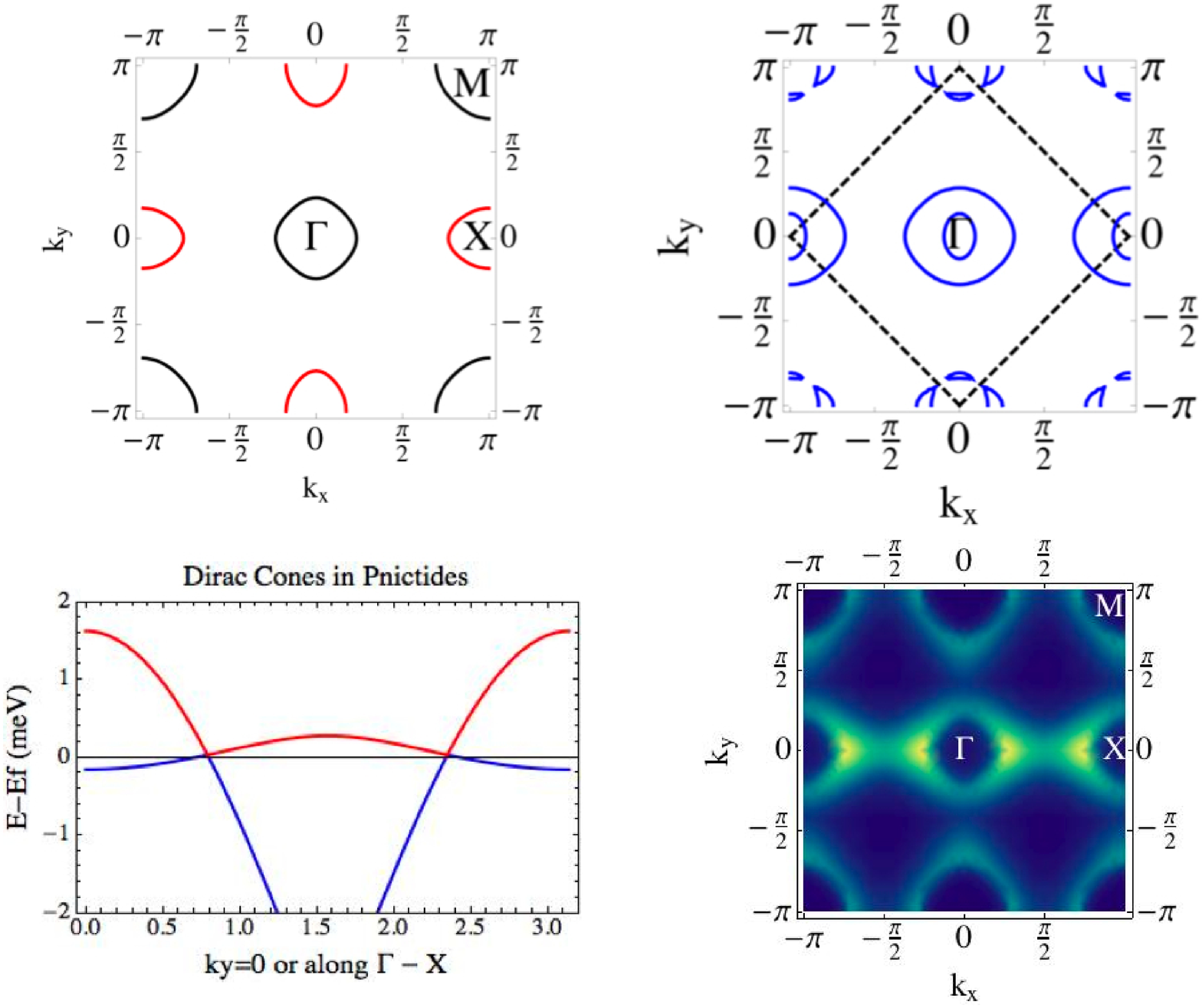}
\caption{(Color online): a) (top left) Unfolded two-band Fermi surface; b) (top right)
Folded four-band Fermi surface, resulting from the staggered buckling of As ions;
c) (bottom left) Dirac cones along the $\Gamma-X$ direction, after folding of the 
Brillouin zone due to collinear AF order; d) (bottom right) Small hole pockets 
originating from the top of a Dirac cone.}
\label{Fig-Dirac-Cones}
\end{figure}

Next we analyze the quasiparticle excitation spectrum. The introduction of
itinerant carriers into the problem presents us with an important question: 
{\it if and how} the collinear AF order modify the quasiparticle spectrum. 
To investigate this problem we adopt a minimal two-band model for the
quasiparticles, which can be described by the tight binding Hamiltonian 
\beq
\hat{H}_{qp}=\sum_{\langle i,j\rangle,\alpha\beta} t^{\alpha\beta}_{ij}\; 
(d^{\alpha}_{i,\sigma})^{\dag}(d^{\beta}_{j,\sigma})+h.c.,
\eeq
where $(d^{\alpha}_{i,\sigma})^{\dag}$ creates an electron/hole at the site $i$, 
with orbital character $\alpha=d_{xz},d_{yz}$ and spin projection $\sigma$. The
values of the four hopping amplitudes, two for direct $d_{xz,yz}-d_{xz,yz}$ 
hopping and two for crossed $d_{xz,yz}-d_{yz-xz}$, and $d_{yz,xz}-d_{xz,yz}$, 
with different signs, and are the same as used in \cite{Raghu}. Finally, the 
coupling between iron itinerant carriers to the iron local moments is done via 
\cite{Philip-Phillips}
\beq
\hat{H}_{Hund}=-\frac{J_H}{2}\sum_{i,\alpha}(d^{\alpha}_{i,\sigma})^{\dag}\vec{\tau}
(d^{\alpha}_{i,\sigma})\cdot\hat{\bf S}_i,
\label{Hund-Hamiltonian}
\eeq
where $\vec{\tau}$ are the Pauli matrices and $J_H$ is the ferromagnetic Hund's 
exchange coupling constant. 

In order to calculate the quasiparticle spectrum we make use of a semiclassical 
approximation. We split the sum over lattice sites $i$ in (\ref{Hund-Hamiltonian}) 
into sums over sublattices $i_A$ and $i_B$, and we then replace $\hat{\bf S}_{A,B}$ 
by their static equilibrium configuration within the colinear AF ordered state. The 
total quasiparticle Hamiltonian $\hat{H}_{qp}+\hat{H}_{Hund}$ is then quadratic 
in the $(d^{\alpha}_{i,\sigma})^{\dag},(d^{\alpha}_{i,\sigma})$ operators and can 
be easily diagonalized. The eigenvalues are plotted in Fig.\ \ref{Fig-Dirac-Cones}.
In the high temperature magnetically disordered phase the sublattice magnetization 
is zero and the dispersion is the one of the tight binding Hamiltonian $\hat{H}_{qp}$ 
alone, where quasiparticles are characterized by a multi-band Fermi surface, with 
a rather large density of states, Figs.\ \ref{Fig-Dirac-Cones}a) and b). Upon collinear 
AF ordering, however, the doubling of the magnetic unit cell, and the consequent 
Brillouin zone folding, produces slightly anisotropic, hole-like Dirac cones at very 
specific locations, $(\pm 0.25 \pi, 0)$ and $(\pm 0.75 \pi, 0)$, related to the collinear 
structure of the AF order, see Figs.\ \ref{Fig-Dirac-Cones}c) and d), thus contributing 
a very small density of states. These two features are consistent with both recent 
ARPES results, see Fig. 1 of Ref. \cite{ARPES}, and optical spectroscopy \cite{Hall-Spectroscopy}, 
which reveal a strong reduction of the density of states upon AF ordering, for $T<T_{SDW}$ 
(where $T_{SDW}$ is the spin-density-wave transition temperature). The existence 
of Dirac cones in iron pnictides had already been discussed in the literature 
\cite{Cones-Itinerant} by using an itinerant-only model. Here, we have demonstrated 
that, also within a model where local moments interact with a bath of quasiparticles, 
such cones appear at the precise positions observed in experiments. Finally, let us 
remark that, from the point of view of the magnetic spectrum, the bath of carriers 
simply provide a width (finite magnon lifetime) to the magnetic spectral lines, without 
otherwise modifying the pole structure obtained in this work.

We have seen that the collinear AF ground state of the spin-Hamiltonian
(\ref{J1-J2-Hamiltonian}), containing two coupled, interpenetrating N\' eel states, 
gives rise to a second, optical spin-wave branch in the magnetic excitation spectrum, 
that can be identified as an additional step in the integrated neutron intensity in 
SeFe$_{2}$As$_{2}$ \cite{Dai}. It remains to determine the specific polarization of 
each of the two spin-wave branches, through polarized neutron scattering, and to 
look for the effects of such additional optical spin-wave branch in the thermodynamic 
properties of the SrFe$_{2}$As$_{2}$ system, such as the magnetic contribution to 
the specific heat \cite{Next}. We have also seen that, when coupled to a bath of
itinerant quasiparticles, the collinear AF order produces hole-like, slightly anisotropic
Dirac cones as recently observed by ARPES in Ref.\ \cite{ARPES}. 

The authors acknowledge invaluable discussions with A. H. Castro Neto
and Mohammed El Massalami. This work was  supported by CNPq and FAPERJ.

\end{document}